# A Millimeter-wave Technique for correlation and beam combination for cosmology

Siddharth Savyasachi Malu, Indian Institute of Technology Indore
Simrol, Khandwa Road, Indore 453 552, India
Tel: +91 750 980 4944; E-mail: siddharth@iiti.ac.in

*Abstract*- We discuss new techniques and ideas in mm-wave instrumentation that can be used in CMB (Cosmic Microwave background) polarization experiments. Novel techniques in antenna receiver, beam combining and detector systems have resulted in greatly improved sensitivities. We present a few promising approaches and discuss briefly plans for feasibility studies for detecting CMB polarization foregrounds and signal.

*Index Terms*-Cosmology, mm-wave instrumentation, CMB, antenna technology, Fizeau beam combination, quasi-optical techniques

## I. INTRODUCTION

Detection of so-called 'B-modes' in CMB polarization is the biggest challenge in cosmology at present [3,1]. The presence of these B-modes (curl patterns in polarization maps) will confirm the inflationary paradigm; their absence would force us to rethink the entire field of early universe cosmology.

Anisotropies in the CMB were detected by COBE in the late-1980s [10] and CMB polarization was detected for the first time by DASI (an interferometer) in 2001 [2].

While most CMB experiments have been imagers, a few interferometric approaches have been quite successful (DASI, CBI etc.). Also, we know that the B-mode signal is $< 0.1\mu K$, so that its detection will involve exquisite control of systematic effects. Hu et al [11] have shown that various systematic effects in imagers can lead to "leakages" from the 2.7K CMB temperature background to B-modes – this requires a systematic error control of 1 in $10^{10}$ for an accuracy of ~5-10%! On the other hand, interferometers (since they make instantaneous differencing measurements) are not sensitive to the background, but only to $\Delta T$, i.e. anisotropies in temperature, which are $< 100\mu K$, thus requiring an accuracy of 1 in $10^5$ to $10^6$ for the same 5-10% accuracy - difficult, but achievable with current technology.

This stringent systematic error control is the most crucial condition an instrument has to satisfy in order to be able to detect B-modes as low as ~$0.005\mu K$. The other (equally important) condition is that it has to be sensitive enough. While it is relatively easy (compared to imagers) to design a low-systematics interferometer; it is difficult to do so without losing on sensitivity. This is the reason a new approach in CMB interferometry was needed.

In this paper, we present a new experiment meant for the detection and characterization of the Cosmic Microwave Background (CMB) polarization. Our collaboration, called Q and U Bolometric Interferometer (QUBIC henceforth) is a planned ground-based experiment. As its name suggests, QUBIC will use a new idea called 'Bolometric Interferometry' to detect and characterize CMB polarization.

## II. BOLOMETRIC INTERFEROMETRY

As discussed above, the two crucial requirements for a B-mode instrument are:
1. Systematics control
2. Sensitivity comparable to imager of similar size

Imagine a large dish in an imager, divided into several constituent parts, each part able to receive radiation from the sky (receivers can be mirror parts or antennae). Received signal is phase modulated in a controlled way such that one pair of antennae have a unique phase difference. If the fields at the jth and kth antennae are $\mathbf{E_j} = E_j \exp(i[\phi_j(t)+\phi_{gj}])$ and $\mathbf{E_k} = E_k \exp(i[\phi_j(t)+\phi_{gk}])$ respectively, then the output at the detectors (bolometers in QUBIC) is

$$Z_m = |E_j|^2 + |E_k|^2 + 2\mathbf{Re}(\exp i[\Delta\phi_{jk}(t) + \Delta\phi_{g,jk}]) \quad (1)$$

where
$\Delta\phi_{jk}(t)$ = phase modulation for the jk-pair
$\Delta\phi_{g,jk}$ = geometric phase difference, unique to every pair of antennae

By demodulation, only the last part in (1) is extracted, and this part represents interference between signals from the two antennae. Now, interfered signal from every pair of antennae is one point in the fourier transform of the image on the sky (this fact is known as the van Cittert-Zernicke theorem in astrophysics) and is called a "visibility". Every visibility then corresponds to a certain baseline length (distance between 2 antennae in a pair), and therefore a certain angular resolution and multipole moment. The correlation between visibilities yields the power spectrum of CMB fluctuations.

If each antenna is dual-polarization-capable, then we can get visibilities of the Stokes parameters $S=\{I,Q,U,V\}$ which are $N_h(N_h-1)/2$ in number, where $N_h$ is the number of antennae.

We describe a few instrumental design and implementation ideas in QUBIC below.

## III. OVERVIEW

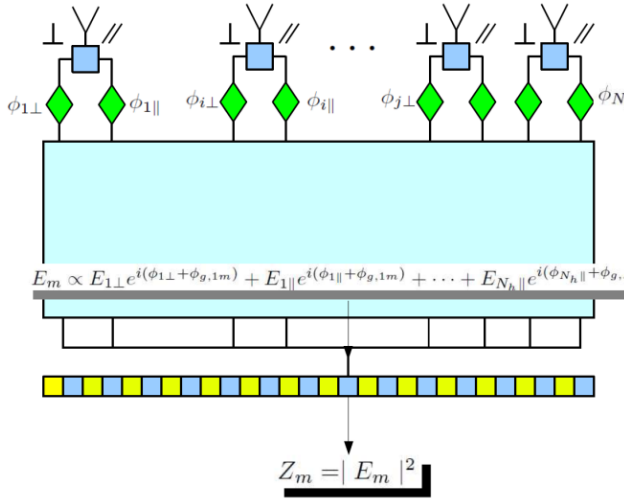

Fig.1 Schematics of QUBIC (figure courtesy Andrea Tartari, APC, Paris)

An overview of the planned instrument is shown in Fig. 1. Signal from horn antennae are split into two polarization states using an OMT (represented as small blue rectangles in the figure), represented in the figure by _ and ∥ - so that for $N_h$ horns, we get $2N_h$ inputs, each of which is phase shifted with the modulators shown as green rhombi. The beam combiner is represented by a large light-blue rectangle and is a Fizeau combiner, i.e. it combines all the $2N_h$ electric fields such that the field from each horn appears at each detector (bolometer). This is achieved geometrically by merely re-radiating from another set of inward-facing antennae, each one of these forming a back-to-back configuration with each sky-facing (input) antenna. Interference or correlation is achieved as described in the previous section.

The first prototype will operate in the W-band with a centre frequency of 93 GHz and will have a 10x10 array of input antennae. The resolution of baselines will range from ~1° to 10°. The next version will have 10x10 antennae in several frequency modules – this is crucial for characterization of foregrounds (for the CMB, anything that exhibits a different spectrum than its well-known and characterized black-body spectrum is a foreground).

## IV. SUBSYSTEMS: NEW IDEAS AND TECHNOLOGY

### A. Sub-band splitting

Visibility is complex quantity, and detectors measure only its real part. In order to extract the imaginary part as well, the standard technique employed in traditional interferometers is to introduce two known phase shifts and then measure the two phase shifted visibilities. In our case, these phase shifts are already provided by the geometry of the Fizeau combiner. Therefore, recovery of the full complex visibility requires just two samplings, i.e. two detectors on the focal plane. However, there will be an array of at least 100 detectors on the focal plane. These can be then used to improve the signal-to-noise (S/N) ratio. However, the extra information can also be represented as visibilities (or signal) in sub-bands, keeping the S/N ratio the same. The amount of information extracted remains the same; its representation changes. QUBIC is unique in this respect: its design allows flexibility in choosing a S/N ratio and frequency resolution – traditional instruments (whether imagers or interferometers) do not have this feature.

The idea that this could be achieved was introduced in [13] for the first time, and Charlassier et al show in [12] that the implementation in simulation works as expected.

*B. Beam Combiner*

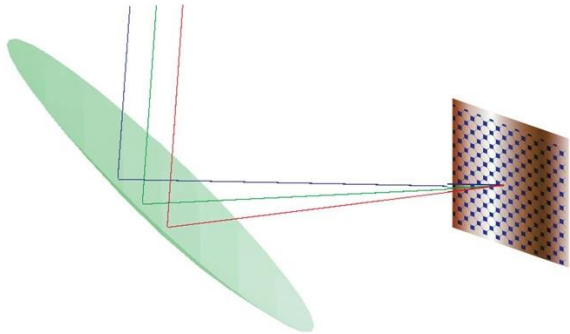

Fig.2 Schematic of the Quasi-optical Fizeau Beam Combiner to be used in QUBIC. The three rays originate from different inward-facing (secondary) horn antennae. These are focused on the detector array on the right, which images the interference pattern (figure courtesy Andrea Tartari, APC, Paris).

Unlike traditional interferometers, where signal is combined pairwise, in a bolometric (or adding) interferometer, signals are added and then detected. As discussed earlier, the Fizeau beam combiner [6,9] adds all the signals such that electric field from every antenna reaches every detector. In a quasi-optical implementation of the Fizeau combiner, this addition is done geometrically – all that is needed is a pair of carefully designed mirrors with minimized aberrations. Off-axis Gregorian systems have these properties, and are therefore the configuration of choice for QUBIC.

## V. CONCLUSIONS

We have presented a novel idea in astronomical interferometry – Bolometric Interferometry. We have also presented two new paradigms in interferometric instrumentation, with greatly enhanced capabilities compared to traditional systems, without compromising on sensitivity and systematic error control. These new ideas present exciting new opportunities as well as unique challenges which are being successfully addressed by our collaboration, towards the construction of a unique instrument capable of testing the most fundamental paradigms behind our current understanding of the early universe.